\documentclass[reprint,amsmath,amssymb,aps,prl]{revtex4-2}
\usepackage{graphicx} % Required for inserting images
\usepackage{amsmath,amssymb}
\usepackage{dsfont}
\usepackage{amsthm}

\usepackage{xcolor}
\usepackage{graphicx}% Include figure files
\usepackage{dcolumn}% Align table columns on decimal point
\usepackage{bm}% bold math
\usepackage{orcidlink}

\usepackage[normalem]{ulem}

\begin{document}

\title{The Lee-Yang property of the Blume-Capel model}

\author{
	{Yuri Kozitsky}\orcidlink{0000-0002-4320-8835}$^1$, 
	{Yurij Holovatch}\orcidlink{0000-0002-1125-2532}$^{2,3,4,5}$
} 
\affiliation{ $^1${Institute of Informatics and Mathematics, Maria Curie-Sk\l odowska University}, {20-031} {Lublin}, {Poland}\\
	$^2${Yukhnovskii Institute for Condensed Matter Physics of the National Academy of Sciences of Ukraine}, {Lviv}, {79011}, {Ukraine}\\
	$^3${$\mathbb{L}^4$ Collaboration and Doctoral College for the Statistical Physics of Complex Systems}, {Lviv-Leipzig-Lorraine-Coventry}, {Europe}\\
		$^4${Centre for Fluid and Complex Systems, Coventry University}, {Coventry} {CV1 5FB}, {UK}\\
	$^5${Complexity Science Hub}, {1030} {Vienna}, {Austria}
}

%\pacs{}
%\date{\today}

\begin{abstract}
The Lee-Yang theory is based on the theorem concerning a property of the ferromagnetic Ising model partition function proved by T. D. Lee and C. N. Yang in 1952. It provides powerful tools for understanding the very nature of phase transitions and critical phenomena in various spin and similar models. 
Sometimes, this theory is applied to models for which the theorem's validity has not been proved. By the main result of E. H. Lieb and A. D. Sokal, Commun. Math. Phys. {\bf 80}, 153 (1981), for a given ferromagnetic model, this validity is guaranteed by the same property of the single-spin partition function. 
Due to the anisotropy term $\sum_i \Delta S_i^2$, the single-spin partition function of the Blume-Capel model fails to have the Lee-Yang property for $\beta \Delta> \ln 2$. In this article, we show that the ferromagnetic interaction in such a model can induce the Lee-Yang property, even for these values of $\beta \Delta$. To the best of our knowledge, it is the first result of this kind.    
\end{abstract}
\maketitle 

\section{Introduction}
%\subsection{Posing}
The Blume-Capel (BC) model \cite{Blume66,Capel66} is a particular case of the Blume-Emery-Griffiths (BEG) model, introduced in \cite{Blume71} to describe the $\lambda$ transition and phase separation in ${\rm He}^3-{\rm He}^4$ mixtures. Both are extensions of the $S=1$ Ising spin model. The Hamiltonian of the Blume-Capel model reads
\begin{equation}
\label{1}
H = - \sum_{\langle i,j\rangle } J_{ij} S_iS_j+ \sum_i \Delta S_i^2,
\end{equation}
with the spin variables $S_i$ taking values $-1, 0,
+1$. As usually, $J_{ij}\geq 0$ is the ferromagnetic exchange interaction and $\Delta$ is
the single-ion anisotropy field. The summations in (\ref{1}) are taken over 
the pairs of sites and over the sites of a crystal lattice, respectively. Typically, this is a
$d$-dimensional 
hyper-cubic lattice $\mathds{Z}^d$. In this case, $i$ is a vector $(i_1, \dots, i_d)$ with integer components. However, other discrete infinite structures can be used as the underlying sets 
of such models.

\begin{figure}[ht!]
\centering
\includegraphics[width=0.9\linewidth]{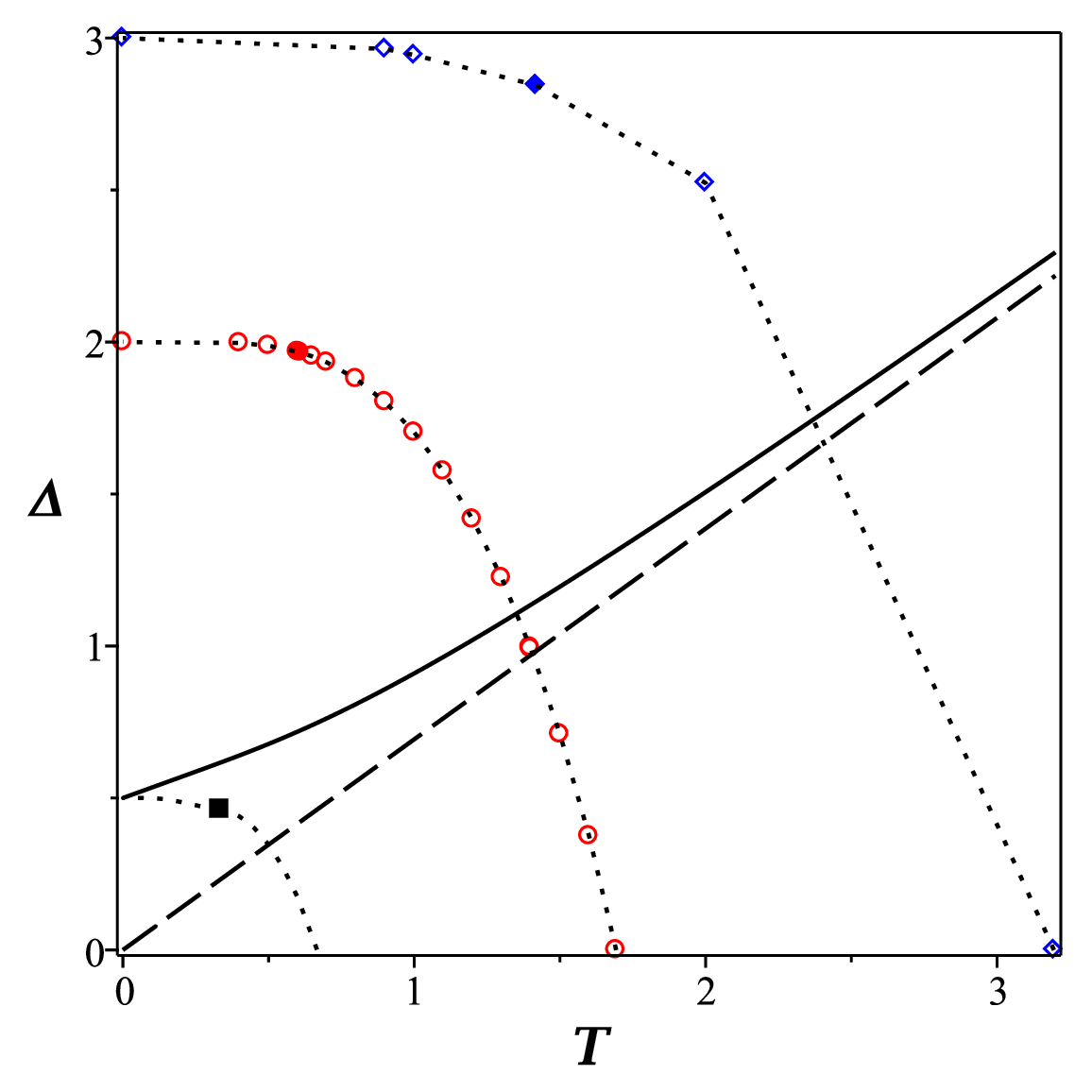}
	\caption{Phase diagram of the Blume-Capel model in the  
    crystal field -- temperature plane at $J=1$: on
		the complete graph and on the $d=2,3$ simple cubic lattices.
		The dotted lines represent phase transition lines; from bottom to top: complete graph, $d=2$, $d=3$, 
		respectively. For the complete graph, the exact result \cite{Honchar25} is shown. 
		For the lattices, the lines serve as a guide. Symbols show the results of 
		the MC simulations (hollow disks for $d=2$ \cite{Zierenberg17} and hollow diamonds for $d=3$
		\cite{Zierenberg15}). 
		Filled symbols show the location of the tricritical point where the 1st order transition line 
        (left from the tricritical point) meets the 2nd order transition line (right from the point).
	The solid and the dashed black lines correspond to the equality in (\ref{7}) and to $\theta=1$, respectively.}
		
\label{fig1}
\end{figure}

Both BC and BEG models, including their quantum extensions \cite{torrico2026}, 
have rich phase diagrams, studied in \cite{Capel66,Blume71,Silva06,Malakis10,Moueddene24c,Moueddene24b,Honchar25,Almeida05,Ghulghazaryan07,jeon2026cluster,
mataragkas2025tricriticality}, and in many other works, see also Fig. \ref{fig1}. The version of \eqref{1} with $\Delta =0$ and $J_{ij}\geq 0$ has the Lee-Yang property for all spin values $S$: for an arbitrary finite portion of the lattice sites $\Lambda$, all zeros of the function 
\begin{equation}
\label{2}    
\Xi_\Lambda (z) = {\rm Tr}\left[ z^{\sum_{i\in \Lambda} S_i }\exp\left(-\beta H_\Lambda \right)\right],
\end{equation}
lie on the unit circle in the complex plane. Here $H_\Lambda$ is the Hamiltonian obtained from that in \eqref{1} by restricting the sums to run over $\langle i,j \rangle \subset \Lambda$ and $i\in \Lambda$, respectively, and Tr denotes the summation over all relevant spin variables.
For $S=1/2$, this property of $\Xi_\Lambda$ is exactly the celebrated Lee-Yang theorem; its extension to all  $S>1/2$ is explained below. 
Observing the evolution of the spin of a quantum probe in a bath \cite{Wei12}, the validity of the Lee-Yang theorem was experimentally confirmed; see \cite{Peng15,gao2024}. 

The Lee-Yang theory based on this result -- as well as its extensions to multicomponent spin and lattice field models, see \cite{dunlop1975multicomponent,kozitsky1997hierarchical,kozitsky2026lee} and \cite[Chapter 2]{albeverio2009statistical} -- proved to be seminal for both theoretical physics  \cite{Ruelle69,Glimm1987,Lieb81,Biskup00,Bena05,frohlich2012some,Basar21,marchetti2024,Simon26} and pure mathematics \cite{Borcea09a,Borcea09b,Branden20}. Significantly, the Lee-Yang theory has deep intrinsic connections with the celebrated Riemann problem and efforts to resolve it; see \cite{newman1976,wolf2020will,zhang2026equivalence}.  
In theoretical physics, the Lee-Yang theory provides a powerful tool for studying phase transitions and critical phenomena \cite{Ruelle69,Biskup00,Bena05,frohlich2012some,deger2020lee}, applicable to models that have the Lee-Yang property. At the same time, the validity of the Lee-Yang theorem for the BC model for all values of $\Delta$ has not  been established, a fact that is often not taken into account by physicists who deal with this model. The purpose of the present work is to contribute to the development of this issue.

The Lee-Yang property of the model \eqref{1} can also be formulated as the statement that all zeros of 
the partition function, cf. \eqref{2},
\begin{equation}
\label{3}    
Z_\Lambda (h) = \Xi_\Lambda (e^h)
\end{equation}
lie on the imaginary axis of the complex plane. By \eqref{1} -- \eqref{3}, one can write 
\begin{gather}
\label{4}
Z_\Lambda (h) = \exp\left(\sum_{\langle i,j \rangle \subset \Lambda} \beta J_{ij} D_iD_j\right)\prod_{ i\in \Lambda} Z_1 (x_i)\bigg{|}_{x_i = h},
\end{gather}
where $D_i = \partial /\partial x_i$ and 
\begin{equation}
\label{5}
Z_1 (x) = {\rm Tr} e^{-\beta\Delta S^2 + x S} = 1 + 2 e^{-\beta \Delta}\cosh x.
\end{equation}
 By the Lieb-Sokal theorem \cite{Lieb81}, see also below, the function $Z_\Lambda$  has the property in question if the function in \eqref{5} does. At the same time, cf. \eqref{3}, by \eqref{5} we have
\begin{eqnarray}
\label{6}
\Xi_1 (z) & = & {\rm Tr}\left[z^S e^{-\beta \Delta S^2}\right]   = 1+ e^{-\beta \Delta} (z+ z^{-1})\\[.2cm] \nonumber & = & e^{-\beta \Delta} z^{-1} P_\theta (z) = e^{-\beta \Delta} z^{-1}(z^2 + 2\theta z +1),
\end{eqnarray}
where $\theta = e^{\beta \Delta}/2$. Then the zeros of $\Xi_1$ are 
\begin{equation}
\label{z}
z_{\pm } = - \theta \pm \sqrt{\theta^2 - 1}.
\end{equation}
For $\theta\leq 1$, they satisfy $|z_{\pm}| =1$, and thus the zeros of $Z_1$ are purely imaginary.
For $J_{ij}\geq 0$, according to the Lieb-Sokal theorem, this implies that also $Z_\Lambda$ has purely imaginary zeros for such values of $\theta$ and an arbitrary $\Lambda$. With $\theta =1/2$, this includes the case of the $S=1$ ferromagnetic Ising model mentioned above. The validity of such a statement for the ferromagnetic Ising model with all standard values of $S$ was proved in \cite{griffiths1969}. 

For $\theta >1$, both $z_{\pm}$ in \eqref{z} are real and lie outside the unit circle. This means that the Lieb-Sokal theorem, or any similar statement, no longer guarantees the desired property of $ Z_\Lambda$. In the Lieb-Sokal theorem, the requirement that $Z_1$ has only imaginary zeros is not a necessary condition. Hence, there may be the possibility that the application of the differential operator in \eqref{4} results in the corresponding property of $Z_\Lambda$ even if $Z_1$ does not have it. In this article, we show that such a possibility does exist. In particular, by our result, the following is true. Consider the model \eqref{1} defined on the lattice $\mathds{Z}^d$ with a nearest-neighbor ferromagnetic interaction:  $J_{ij}=J>0$ for the pairs $\langle i,j \rangle$ satisfying $|i-j|=1$, and $J_{ij}=0$ otherwise. Here $|\cdot |$ stands for the Euclidean distance. For an arbitrary integer $L$, let $V_L$ be a parallelepiped subset of the lattice $\mathds{Z}^d$ consisting of $2L$-long strings of neighboring sites parallel to one of the axes, and let $\Lambda = V_L$, i.e., $Z_{V_L}$ be the partition function of this model as in \eqref{4}, \eqref{5}. Then all zeros of $Z_{V_L}$ are purely imaginary whenever $\beta$, $J$, and $\Delta$ satisfy the following condition
\begin{eqnarray}
\label{7}
e^{2\beta \Delta}/4  \leq  \cosh (\beta J).
\end{eqnarray}
This means that $Z_{V_L}$ has the property in question for every $\Delta$ and sufficiently large $J$. For a fixed $J$, the area of validity of \eqref{7} is illustrated in Fig. 1. 

In the following sections, we formulate, prove, and discuss our (mathematical) statement and its applications to the physical models described by \eqref{1}. To the best of our knowledge, this is the first rigorous demonstration that a ferromagnetic interaction can induce the Lee-Yang property.

\section{The Statement: Proof and Applications}
Except for the Lieb-Sokal theorem \cite{Lieb81} and the corresponding notions from complex analysis - which we formulate here in an accessible form -
the mathematics used below amounts to elementary algebraic operations with complex numbers, the set of which is denoted by $\mathds{C}$. By $\Re x$ and $|x|$ we denote the real part and the modulus of $x\in \mathds{C}$, respectively. By $\mathds{C}^N$ we mean the space of complex vectors $(x_1, \dots, x_N)$. 

\subsection{Preliminaries and the statement}

A function, $\Phi$, of $N$ complex variables is called \emph{entire} if its Taylor expansion around any $(x^*_1, \dots , x^*_N)$ has a nonzero radius of convergence; see \cite{levin1996lectures}. Such a function is \emph{of exponential type} if there exist positive $C$ and $\rho$ such that the following estimate
\[
|\Phi(x_1, \dots , x_N)| \leq C \exp\left[\rho (|x_1| + \cdots +|x_N| )\right]
\]
holds everywhere on $\mathds{C}^N$. In particular, all complex polynomials satisfy both of these conditions.
The principal example of such a function is 
\begin{equation}
\label{7a}
\widehat{\Phi}(x_1, \dots , x_N) = \prod_{l=1}^N Z_1 (x_l),
\end{equation}
where $Z_1$ is as in \eqref{5}. Note that $\cosh x$ is an exponential-type entire function of $x\in \mathds{C}$. 

An exponential-type entire function, $\Phi$, is called \emph{stable},  cf. \cite{Borcea09a}, if
$\Phi(x_1, \dots, x_N)\neq 0$ whenever $\Re x_l >0$ for all $l$. By \eqref{6} and \eqref{z}, it follows that the function in \eqref{7a} is stable only for $e^{\beta \Delta}/2=\theta \leq 1$. 
If $\Phi$ is stable and symmetric, i.e., is such that $\Phi(x_1, \dots, x_N) = \Phi(-x_1, \dots, -x_N)$, then the function $\varphi(x)= \Phi(x, \dots, x)$ may have only imaginary zeros. By the latter fact, stable symmetric functions are suitable for proving the Lee-Yang property.  
A paramount problem of the theory of stable functions is to characterize operators that preserve their stability; see \cite{Borcea09a,Borcea09b,Branden20}. A particular solution of this problem is provided by the Lieb-Sokal theorem \cite[Proposition 2.2]{Lieb81}, which we formulate here in the form adapted to the context. Consider
\begin{eqnarray}
\label{8}
& & F(x_1, \dots , x_N)\\ & & = \exp\left(\frac{1}{2} \sum_{1\leq l,l' \leq N} K_{l,l'} D_l D_{l'} \right) \Phi (x_1, \dots , x_N), \nonumber
\end{eqnarray}
where, similarly to \eqref{4}, $D_l$ stands for $\partial / \partial x_l$ and $\Phi$ is an exponential-type entire function. One observes that here $F$ is obtained from $\Phi$ by means of an infinite-order differential operator, the action of which should be carefully defined. By means of methods developed in \cite{kozitsky2003infinite}, it is possible to show that $F$ in \eqref{8} is an exponential-type entire function if $\Phi$ is such a function. The key issue is the stability of $F$.

According to the Lieb-Sokal theorem, the following is true: if $\Phi$ in \eqref{8}
is stable, then $F$ is also stable whenever $K_{l,l'}\geq 0 $ for all $l,l'$. 

Using in \eqref{8}  $\Phi= \widehat{\Phi}$ given in \eqref{7a} and \eqref{5}, we conclude that the corresponding $F$ is stable and symmetric; hence, 
$Z_\Lambda (h) = F(h, h, \dots , h)$
has only imaginary zeros for $\theta \leq 1$ and an arbitrary portion $\Lambda$ of $N$ spins. Note that a stable exponential-type entire function, $\phi$, can avoid zero, i.e., $\phi(x) \neq 0$ for all $x\in \mathds{C}$. An example can be $\phi(x) = e^x$. The fact that $Z_N$ has zeros follows from its evenness $Z_N(h) = Z_N (-h)$ and Hadamard's representation theorem; see \cite{levin1996lectures}.

By means of the Lieb-Sokal theorem, we prove the following statement.

{\bf Statement:} \emph{Let the entries of \eqref{8} have the following properties: (a) $N$ is even, i.e., $N=2M$; (b) $\Phi= \widehat{\Phi}$ given in \eqref{7a} and \eqref{5}; (c) $K_{l,l'} \geq 0$ for all $l,l'$, and there exists a division of the set of indices $1,2, \dots, 2M$ into disjoint pairs (dimers) $\ell_k= \{l_k, l'_k\}$, $k=1,2, \dots, M$ such that} 
\begin{equation}
\label{Ne}
K_{l_k, l'_k} \geq \kappa \ \ {\rm for} \ {\rm some} \ \kappa>0 \ {\rm and} \ {\rm all} \ k.
\end{equation} 
\emph{Then the corresponding function $F$ is stable if the following condition is satisfied}
\begin{equation}
 \label{Ne1}
 e^{2 \beta \Delta}/4 = \theta^2 \leq \cosh \kappa.
\end{equation}

Let us make a few preliminary comments on this statement. 
A detailed analysis and discussion will follow in the next section.
By \eqref{Ne1}, it is immediate that our statement is an extension to $\theta>1$ of the corresponding direct outcome of the Lieb-Sokal theorem, which was mentioned above. The price of this extension is not too high -- the evenness of $N$ and the additional positivity of the interaction matrix $(K_{l,l'})$ expressed in item (c) and \eqref{Ne}. Clearly, standard examples of such matrices, including the one mentioned above, cf. \eqref{7}, have this property. Below, we discuss this issue in more detail. At the same time, \eqref{Ne1} is again a sufficient condition -- its violation does not yet imply the lack of the Lee-Yang property.

\subsection{The proof}

Naturally, we can limit ourselves to considering the case of $\theta>1$. Fix such a $\theta$ and pick positive $\varkappa \leq \kappa$ that verifies, cf. \eqref{Ne1}, 
\begin{equation}
\label{9} 
\cosh \varkappa = \theta^2.
\end{equation}
Then we split $K_{l,l'} = K'_{l,l'} + K''_{l,l'}$ according to
\begin{eqnarray}
\label{Ne2}
& & K'_{l_k,l'_k} = \varkappa, \ \   K''_{l_k,l'_k} = K_{l_k,l'_k} - \varkappa,  \ {\rm for}  \ {\rm each} \ {\rm dimer} \  \ell_k, \nonumber \\  & & K'_{l,l'} = 0, \ \ K''_{l,l'} = K_{l,l'}, \quad {\rm otherwise}.
\end{eqnarray}
In view of condition (c) of the statement, we then have
\begin{equation}
\label{Ne3}
K'_{l,l'} \geq 0 \ \ {\rm and} \ \ K''_{l,l'} \geq 0 \ \ {\rm for} \ {\rm all} \ l,l'. 
\end{equation}
Now we introduce the function, cf. \eqref{7a} and \eqref{8},
\begin{eqnarray}
\label{10}
& & G (x_1, \dots , x_{2M}) \\ \nonumber & & \ \  =  \exp\left(\frac{1}{2} \sum_{1\leq l,l' \leq 2M}K'_{l,l'}D_i D_j\right) \prod_{1\leq l\leq 2M} Z_{1}(x_l) \\[.2cm] \nonumber& & \ \ =  \exp\left( \varkappa\sum_{k=1}^M D_{l_k} D_{l'_k}\right)\prod_{1\leq l\leq 2M} Z_{1}(x_i) \\[.2cm] \nonumber & & \ \ =  \prod_{k=1}^M \bigg{[}e^{\varkappa D_{l_k} D_{l'_k} } Z_1 (x_{l_k})Z_1(x_{l'_k}) \bigg{]} , 
\end{eqnarray}
for which by \eqref{8} and \eqref{Ne2} we get
\begin{eqnarray}
\label{11}
& & F(x_1, \dots , x_{2M}) \\ \nonumber & &  =  \exp\left(\frac{1}{2} \sum_{1\leq l,l' \leq 2M}K''_{l,l'}D_l D_{l'}\right) G (x_1, \dots , x_{2M}).
\end{eqnarray}
Then the stability of $F$ will follow by the Lieb-Sokal theorem and the positivity in \eqref{Ne3}, provided that $G$ given in \eqref{10} is stable. By \eqref{10},\eqref{5}, and \eqref{6}, we get
\begin{equation}
\label{12a}   
G (x_1, \dots , x_{2M})  =  \prod_{k=1}^M \Psi (x_{l_k}, x_{l'_k}),
\end{equation}
where
\begin{eqnarray}
\label{13}
& &  \Psi (x, y)   =  e^{\varkappa D_{x} D_{y} } Z_1 (x)Z_1(y) \\ \nonumber & &  \ \ =  {\rm Tr} \exp\left(\varkappa S_1 S_2 - \beta \Delta (S_1^2+S_2^2) + x S_1 + y S_2  \right), 
\end{eqnarray}
with $D_x = \partial/\partial x$ and $D_y = \partial/\partial y$.
Thus, $G$ is stable if $\Psi$ is so. To prove the latter, we calculate the trace in \eqref{13}
and obtain
\begin{eqnarray*}
 \Psi (x,y)  &  = &  4e^{-2\beta \Delta}\bigg{[} \frac{e^\varkappa}{2} \cosh (x+y) \\ & + &\frac{e^{-\varkappa}}{2} \cosh (x-y) \\ & + & \theta (\cosh x + \cosh y) + \theta^2 \bigg{]},  
\end{eqnarray*}
 where $\theta =  e^{\beta \Delta}/2$, see \eqref{Ne1}. Now we use the identities $\cosh 2\alpha = 2 \cosh^2 \alpha -1$ and $\cosh \alpha + \cosh \alpha' = 2 \cosh (\alpha + \alpha')/2\cosh (\alpha -\alpha')/2$, 
 take into account \eqref{9}, and transform the latter into the following expression
\begin{eqnarray}
 \label{14}
 \Psi (x, y) & = & 4e^{-2\beta \Delta}\left[e^\varkappa c^2_+ + e^{-\varkappa} c^2_{-} + 2 \theta c_+ c_- \right] \\[.2cm] \nonumber
 & = & 4e^{-2\beta \Delta+ \varkappa} (c_{+} + \omega_{+} c_{-}) (c_{+} + \omega_{-} c_{-}),
 \end{eqnarray}
where $c_{\pm} = \cosh[(x \pm y)/2]$,  $\varkappa$ and $\theta$ are subject to \eqref{9}, and
\begin{equation}
\label{15}
\omega_{\pm} = e^{-\varkappa}[\theta \pm \sqrt{\theta^2-1}]. 
\end{equation}
Then the stability of $\Psi (x, y)$ can be proved by showing that both 
\[
\psi_{+}(x, y)=c_{+} + \omega_{+} c_{-}, \   \ \psi_{-}(x, y)=c_{+} + \omega_{-} c_{-}
\]
do not vanish whenever $\Re x>0$ and $\Re y >0$. Let us show that
$\theta>1$ (hence $\varkappa>0$, see \eqref{9}) imply
\begin{equation}
\label{Ne5}
0 < \omega_{-} < \omega_{+}< 1.
\end{equation}
The first two inequalities are immediate; thus, we prove $\omega_{+}< 1$.
By \eqref{9}, it follows that $e^\varkappa = \theta^2 +\sqrt{\theta^4 -1}$. Hence, to get the property in question, we write
\begin{gather*}
e^\varkappa - (\theta + \sqrt{\theta^2-1} ) = \theta (\theta-1) +  \sqrt{\theta^2-1}(\sqrt{\theta^2+1}-1),  
\end{gather*}
which is strictly positive for $\theta>1$. This completes the proof of \eqref{Ne5}, by which
both
\begin{equation}
\label{16}
\varepsilon_{\pm} = \frac{1-\omega_{\pm}}{1+\omega_{\pm}}
\end{equation}
lie in $(0,1)$. At the same time, by means of \eqref{16}, one can write
\begin{equation}
\label{17}
\psi_{\pm} (x , y) = \cosh \frac{x}{2} \cosh \frac{y}{2} + \varepsilon_{\pm} \sinh \frac{x}{2} \sinh \frac{y}{2}.
\end{equation}
If $\Re x>0$ and $\Re y>0$, then $|\cosh (x/2)|>0$, $|\sinh (x/2)|>0$, $|\cosh (y/2)|>0$, $|\sinh (y/2)|>0$. The latter inequalities yield $c e^{i\alpha}:=\tanh (y/2) \neq 0$. Assume now that $\psi_{+} (x , y)=0$ for some $x, y$ such that $\Re x>0$ and $\Re y>0$. By \eqref{17}, this yields
\[
\frac{e^{x}+1}{e^{x}-1}= \frac{\cosh \frac{x}{2}}{\sinh \frac{x}{2}} = - \varepsilon_{+} \frac{\sinh \frac{y}{2}}{\cosh \frac{y}{2}} =  - \varepsilon_{+}ce^{i\alpha}. 
\]
We solve this with respect to $e^{x}$ and thus get
\[
e^{x}= \frac{\varepsilon_{+}c e^{i\alpha }-1}{\varepsilon_{+}c e^{i\alpha }+1}= \frac{\varepsilon_{+}c \cos \alpha -1 + i \varepsilon_{+}c \sin \alpha}{\varepsilon_{+}c \cos \alpha +1 + i \varepsilon_{+}c \sin \alpha} ,
\]
from which we finally arrive at the following formula
\begin{eqnarray}
\label{18}
|e^{x}|^2 =  \frac{1+\varepsilon_{+}^2 c^2 - 2 \varepsilon_{+} c \cos \alpha}{1+\varepsilon_{+}^2 c^2 + 2 \varepsilon_{+} c \cos \alpha}.
\end{eqnarray}
In a similar way, we solve $\tanh \frac{y}{2}= ce^{i\alpha}$ and obtain
\begin{equation}
\label{18a}
 |e^{y}|^2  =  \frac{1+c^2 + 2  c \cos \alpha}{1+ c^2 - 2  c \cos \alpha}.
\end{equation}
For $\cos \alpha \geq 0$, by \eqref{18} one gets $|e^{x}|^2\leq 1$ and $|e^{y}|^2\geq 1$; while, for $\cos \alpha < 0$, one has $|e^{x}|^2> 1$ and $|e^{y}|^2< 1$; see \eqref{18a}. At the same time, $|e^{x}|^2 = e^{2\Re x}> 1$ and $|e^{y}|^2 = e^{2\Re y}> 1$, which contradicts both versions and therefore excludes the possibility of $\psi_{+} (x, y) =0$ for $\Re x >0$ and $\Re y>0$. Obviously, vanishing $\psi_{-} (x , y)$ can be excluded in the same way just by interchanging $\varepsilon$; see \eqref{Ne5} and \eqref{16}. Thus, $\Psi$ in \eqref{14} is stable. Consequently, $G$ in \eqref{12a} is also stable, and the proof of the statement follows by the Lieb-Sokal theorem; see \eqref{11}.   

\subsection{Applications}

Here we discuss possible applications of the just-proved result to the model described by \eqref{1} with several options of the underlying structure and the interaction matrix $(J_{ij})$. Before starting this discussion, we note that by conditions (a) and (c), the graph $(V,E)$, with vertex set $V=\{1,2, \dots, N\}$ and edge set $E=\{ \{l,l'\}: K_{l,l'} \geq \kappa\}$, is \emph{perfectly matchable}; see \cite{bondy1976graph}. At the same time, for $\theta\leq 1$, the Lee-Yang theorem is unconditionally valid for all possible types of underlying structure.  

As mentioned above, the main aim of the Lee-Yang theory is to describe the analytic properties of the free energy obtained in the thermodynamic limit and to reveal their role in possible phase transitions; see \cite{Ruelle69,Biskup00,Bena05,Basar21,Simon26}. This description is based on the formula
\begin{equation}
\label{12}
Z_{\Lambda} (h) = F_{2M} (h, \dots , h),
\end{equation}
holding for each subset $\Lambda =\{i_1, \dots , i_{2M}\}$ containing $2M$ lattice sites.
This formula readily follows by \eqref{11}, \eqref{10}, \eqref{8}, and \eqref{4} with $K_{l,l'}=\beta J_{i_li_{l'}}$.
Then such $Z_{\Lambda}$ has purely imaginary zeros if $F_{2M}$ is stable. In this case, the limiting free energy
\begin{equation}
\label{FE}
f(\beta,h)= - \lim_{n\to \infty}  \frac{1}{2\beta M_n} \ln Z_{\Lambda_n} (h), \quad M_n \to \infty,   
\end{equation}
is analytic in the domains free from the zeros of $Z_{\Lambda_n} (h)$, $2M_n$ being the cardinality of $\Lambda_n$. In particular, it is analytic in $\{h \in \mathds{C}: \Re h \neq 0\}$ if $F_{2M_n}$ is stable for all $n$. Note, however, that for the latter property to persist for all $n$, the lower bound in \eqref{Ne} should be the same for all such $n$.    

For lattice systems, the thermodynamic limit \eqref{FE} is usually taken along an exhausting sequence $\{\Lambda_n\}$ of finite subsets of the underlying discrete structure, e.g., of the hyper-cubic lattice $\mathds{Z}^d$. In the latter case, under a mild (van Howe) condition that excludes exotic examples of such subsets, see \cite[Chapter 2]{Ruelle69}, the limiting free energy $f$ is independent of the particular choice of the sequence $\{\Lambda_n\}$. This gives us the freedom to choose each $\Lambda_n$, consisting of $2M_n$ sites, cf. \eqref{FE}, in such a way that both mentioned conditions (a) and (c) are satisfied for most of the standard examples of interaction matrices. 
In view of this, an option can be the model \eqref{1} defined on $\mathds{Z}^d$ with a nearest-neighbor interaction: $J_{ij}=J$ for $|i-j|=1$, and $J_{ij}=0$ otherwise. Then each $\Lambda_n$ can be taken as a collection of parallel paths of nearest neighbors that contain an even number of sites. In particular, $\Lambda_n$ can be a parallelepiped of length $2L$ along the $OX$ axis. Since an even-length path is perfectly matchable, we can match pairs of neighbors along each such path and set $\kappa= \beta J$, cf. \eqref{7} and \eqref{Ne1}. Clearly, any sequence of such $\Lambda_n$ satisfies the van Howe condition: the ratio (``the number of neighboring sites outside $\Lambda_n$")/(``the number of sites in $\Lambda_n$") tends to zero as $n\to \infty$; see \cite[Chapter 2]{Ruelle69}. The same approach can also be applied if the parallelepipeds $\Lambda_n$ and the matrix $(J_{ij})$ are subject to periodic boundary conditions. For long-range interactions $J_{ij}$, we take $\kappa= \beta \max_{1\leq l \leq d}J_{ii_l}$, $i_l-i = 1_l$, where $1_l= (0, \dots 0, 1,0, \dots 0)$ with $1$ standing on $l$-th position. In this case, $\Lambda_n$ is supposed to be a parallelepiped with an even number of sites in the direction where the maximum is attained. In all these examples, for the parameters that verify  \eqref{7}, the corresponding model has the Lee-Yang property, and hence the limiting free energy $f$ is analytic for $h\neq 0$. The examples just presented are the simplest ones -- if necessary, plenty of more sophisticated versions of such constructions can be realized. This naturally applies to the versions of \eqref{1} based on trees, hierarchical lattices, or other graphs; see \cite{rocha2018,ekiz2023,rocha2023}.   

\section{Discussion and Remarks}

For the Ising model with $S=1/2$, the Lee-Yang theorem implies that zero is the only point on the real line where the limiting free energy can be nonanalytic as a function of $h$. At the same time, this model undergoes only a second-order phase transition, which means that the spontaneous magnetization $M$ is a continuous function of the temperature. For the Blume-Capel model with  the Curie-Weiss interaction \cite{Honchar25}, $M$ is obtained as $y_*/\beta$, where $y_*$ is a point of global maximum of the function
\begin{equation*}
%\label{E}
E(\beta, y)= - y^2/2\beta + \ln Z_1 (y),
\end{equation*}
see \eqref{5}. For $e^{\beta \Delta}/2 = \theta \leq 1$, the zeros of $Z_1$ lie on the imaginary axis. By Hadamard's representation theorem \cite{levin1996lectures}, this implies  
\begin{equation}
\label{E}
E(\beta, y)= - y^2/2\beta +  \ln (1+2 e^{-\beta \Delta})+ \sum_{j=1}^\infty \ln (1 + \gamma_j y^2),
\end{equation}
where positive numbers $\gamma_j\geq \gamma_{j+1}$ satisfy $\sum_j \gamma_j < \infty$, see, e.g., \cite{kozitsky1997hierarchical} for more details. Therefore, $M$ has to be found from the equation
\begin{equation}
 \label{E1}
 M \left( 1- 2\beta \sum_{j=1}^\infty \frac{\gamma_j}{1+ \gamma_j \beta^2 M^2} \right) =0.
\end{equation}
For $\beta \leq \beta_c:= 1/(2 \sum_{j} \gamma_j)$, the only solution of \eqref{E1} is $M=0$. By \eqref{5}, the latter means that $\beta_c$ is the smaller solution of equation $\beta = 1+\theta = 1 + e^{\beta \Delta}/2$, that exists for $\Delta \leq W(2/e)$, which is slightly smaller than $0.5$, cf. Fig.1. Here $W$ is Lambert's function.  
For $\beta>\beta_c$, \eqref{E1} has a positive solution, $M(\beta)$, such that $y_*(\beta) = \beta M(\beta)$ maximizes $E$ in \eqref{E}. By standard arguments, it follows that $M(\beta)$ is a continuous function such that $M(\beta)\sim (\beta - \beta_c)^{1/2}$ as $\beta \downarrow \beta_c$. Thus, in this case, the presence of the Lee-Yang property implies that the phase transition must be of second order. By contraposition, the discontinuity of $M(\beta)$ at $\beta_c$ implies the absence of the Lee-Yang property. Recall that we are talking about the Blume-Capel model with the Curie-Weiss interaction, for which our present result is not applicable, as the corresponding interaction intensities $J_{ij} = J/N$ are not separated from zero, uniformly in $N$.      

The relationship between the Lee-Yang property and the order or type of phase transition has long been, and remains, a subject of enduring interest. Along with the just-mentioned model, numerous examples of models are known that manifest the first-order phase transition (i.e., $M(\beta)$ is discontinuous) where this property is absent; see \cite{Yamada81,Lee94,Biskup00,Ghulghazaryan07}.  

To illustrate our result, let us consider the nearest-neighbor interaction in \eqref{1} and choose the units of $\beta$ such that $\beta = 1/JT$. Then we define $D = \Delta/J$ and set its dependence on $T$ assuming equality in \eqref{7}. This yields the function
\begin{equation}\label{D}
D(T) = \frac{T}{2}\left[\ln \left(e^{1/T} + e^{-1/T} \right)+ \ln 2\right],
\end{equation}
which is plotted as a solid black line in Fig. \ref{fig1}. The dashed line there represents the function $D'(T) = T \ln 2$, obtained by setting $\theta = 1$. According to the Lieb-Sokal theorem, the Lee-Yang property of the model considered takes place for values of $\Delta, \beta, J$ that correspond to points below the dashed line, while our result extends this statement to all points below the solid line. Remarkably, for both $d=2$ and $d=3$, parts of the second-order transition lines — where the Lee-Yang property is expected to hold — lie above the solid line. This indicates that further extensions of our result may be possible. 
The lowest dotted curve in Fig. \ref{fig1} represents the aforementioned result of \cite{Honchar25}, for which the order parameter satisfies \eqref{E1}. As seen in the figure, the entire dotted curve lies in the regions where the nearest-neighbor interacting Blume-Capel spins have the Lee-Yang property. At the same time, the upper part of this curve - that lies above the black square - corresponds to the absence of the Lee-Yang property, which follows from the analysis made at the beginning of this section. This points to the importance of condition (c) of our statement, according to which $J_{ij}$ should be separated from zero uniformly in $N$.

As we have mentioned several times, our statement yields a sufficient condition for the model in question to have the Lee-Yang property. In the case of nearest-neighbor interacting Blume-Capel spins, our result may probably have extension beyond the validity of 
\eqref{7}. However, such an extension would certainly depend on the lattice dimension (and on other details of the interaction matrix $(J_{ij})$). For $d=1$, the corresponding numerical results can be found in \cite{Almeida05,Ghulghazaryan07}. According to \cite[Fig. 1]{Ghulghazaryan07}, the zeros of   
$\Xi_N$, see \eqref{2}, lie on the unit circle for $e^{\beta \Delta} = 3$ and $e^{\beta J}=4$. For these values, we have
${\rm LHS (\eqref{7})} = 18/8$ and ${\rm RHS (\eqref{7})} = 17/8$. Hence, the Lee-Yang property persists even if the upper bound in \eqref{7} is exceeded. At the same time, for $e^{\beta \Delta} = 3$ and $e^{\beta J}=20/9$, the zeros leave the unit circle. In this case, ${\rm LHS (\eqref{7})} = 18/8= 810/360$, while ${\rm RHS (\eqref{7})} = 481/360$. These observations indicate that the sharp bound for $e^{\beta \Delta}$ could be quite close to that given in \eqref{7}.      
We plan to address these issues in a forthcoming article.

\section{acknowledgments}
The authors thank all referees, whose valuable comments and suggestions were essential.   
Yu.H. was supported by the National Research Foundation of Ukraine, Project 2023.03/0099
``Criticality of complex systems: fundamental aspects and applications''. 

\section{Statements}
\begin{itemize}
    \item Data availability: no specific data are related to this research.
    \item Research involving human or animal subject: not applicable.
    \item Dual use research: not applicable.
    \item Use of AI-based writing tools: such tools were not used in writing this article.
\end{itemize}

%\bibliography{references_BC_Lee_Yang}

\end{document}